\documentclass[12pt]{article}
\usepackage{epsf}
\usepackage[dvips]{graphicx,psfrag}

\renewcommand{\thefootnote}{\#\arabic{footnote}}
\begin{document}

\newcommand{\gtrsim}{ \mathop{}_{\textstyle \sim}^{\textstyle >} }
\newcommand{\lesssim}{ \mathop{}_{\textstyle \sim}^{\textstyle <} }

\renewcommand{\thefootnote}{\fnsymbol{footnote}}
\setcounter{footnote}{0}
\begin{titlepage}

\def\thefootnote{\fnsymbol{footnote}}

\begin{center}

\hfill hep-ph/0410287\\
\hfill October, 2004\\

\vskip .4in

{\Large \bf

Hadronic decay of the gravitino in the early universe and its
implications to inflation\footnote{Talk given at PASCOS'04 (Boston,
MA, August 16 -- 22, 2004).}

}

\vskip .4in

{\large
 Masahiro Kawasaki$^{(a)}$,
 Kazunori Kohri$^{(b)}$\footnote{\uppercase{P}resent address:
\uppercase{H}arvard-\uppercase{S}mithsonian \uppercase{C}enter for
\uppercase{A}strophysics, 60 \uppercase{G}arden \uppercase{S}t.,
\uppercase{C}ambridge, \uppercase{MA} 02138, \uppercase{U.S.A.}}
 and
 Takeo Moroi$^{(c)}$
}

\vskip .3in

{\em $^{(a)}$Institute for Cosmic Ray Research,
University of Tokyo\\
Kashiwa, Chiba 277-8582, Japan}

\vskip .1in

{\em $^{(b)}$Department of Earth and Space Science\\
Osaka University, Toyonaka, Osaka 560-0043, Japan}

\vskip .1in

{\em $^{(c)}$ Department of Physics, Tohoku University,  Sendai
980-8578, Japan}

\end{center}

\vskip .4in

\begin{abstract}

We discuss the effects of the gravitino on the big-bang
nucleosynthesis (BBN), paying particular attention to the hadronic
decay mode of the gravitino.  We will see that the hadronic decay of
the gravitino significantly affect the BBN and, for the case where the 
hadronic branching ratio is sizable, very stringent upper bound on the 
reheating temperature after inflation is obtained.

\end{abstract}

\end{titlepage}

\renewcommand{\thepage}{\arabic{page}}
\setcounter{page}{1}
\renewcommand{\thefootnote}{\#\arabic{footnote}}
\setcounter{footnote}{0}

\section{Introduction}

It has long been regarded that the gravitino, superpartner of the
graviton in the supergravity theory, may cause serious problem in
cosmology\cite{Weinberg:zq,Moroi:1995fs}.  This is because the
primordial gravitinos produced in the very early universe decay with
vary long lifetime.  In particular, if the gravitino mass $m_{3/2}$ is
smaller than $\sim O(10\ {\rm TeV})$, its lifetime is expected to
become longer than $1\ {\rm sec}$ so the primordial gravitinos decay
after the big-bang nucleosynthesis (BBN) starts.  Even with the
inflation, such a problem may not be solved; even if the gravitinos
are diluted by the inflation, gravitinos are produced by the
scattering processes of the thermal particles after the reheating.  As
we will discuss later, the gravitino abundance becomes larger as the
reheating temperature becomes higher.  Consequently, in order not to
spoil the success of the BBN scenario, upper bound on the reheating
temperature is obtained, and the primordial gravitinos provide
stringent constraints on cosmological scenarios based on (local)
supersymmetry.

Thus, the effects of the gravitinos on the BBN have been intensively
studied in many studies.  In particular, in the past, the BBN with
hadrodissociation processes induced by hadronic decays of long-lived
particles was studied in several articles\cite{HadronicDecay}.  After
those studies, however, there have been significant theoretical,
experimental and observational progresses in the study of the BBN.
Thus, we performed a new analysis taking account of those progresses,
paying a special attention to the hadronic decay of the
gravitino\cite{Kawasaki:2004qu}.  The most important improvements
compared to the old works are as follows. (i) We carefully take into
account the energy loss processes for high-energy nuclei through the
scattering with background photons or electrons. In particular,
dependence on the cosmic temperature, the initial energies of nuclei,
and the background $^4{\rm He}$ abundance are considered. (ii) We
adopt all the available data of cross sections and transfered energies
of elastic and inelastic hadron-hadron scattering processes. (iii) The
time evolution of the energy distribution functions of high-energy
nuclei are computed with proper energy resolution. (iv) The JETSET 7.4
Monte Carlo event generator\cite{Sjostrand:1994yb} is used to obtain
the initial spectrum of hadrons produced by the decay of
gravitino. (v) The most resent data of observational light element
abundances are adopted. (vi) We estimate uncertainties with Monte
Carlo simulation which includes the experimental errors of the cross
sections and transfered energies, and uncertainty of the baryon to
photon ratio.

\section{Outline of the Analysis}

Let us briefly discuss the outline of our analysis.  In our study,
we first calculate the gravitino abundances as a function of the
reheating temperature, which is defined as $T_{\rm R}\equiv
\left(\frac{10}{g_* \pi^2}M_*^2\Gamma_{\rm inf}^2\right)^{1/4}$, where
$\Gamma_{\rm inf}$ is the decay rate of the inflaton, and $g_*=228.75$
is the effective number of the massless degrees of freedom in the
MSSM.  By numerically solving the Boltzmann equations governing the
evolution of the number density of the gravitino, we found that the
number density of the gravitino $n_{3/2}$ normalized by the entropy
density $s$ is well approximated by
\begin{eqnarray}
    \frac{n_{3/2}}{s} &\simeq& 
    1.9 \times 10^{-12}
    \times \left( \frac{T_{\rm R}}{10^{10}\ {\rm GeV}} \right)
    \nonumber \\ &&
    \left[ 1 
        + 0.045 \ln \left( \frac{T_{\rm R}}{10^{10}\ {\rm GeV}} 
        \right) \right]
    \left[ 1 
        - 0.028 \ln \left( \frac{T_{\rm R}}{10^{10}\ {\rm GeV}} 
        \right) \right].
\end{eqnarray}
Thus, the number density of the primordial gravitino is approximately
proportional to $T_{\rm R}$ and hence, with higher reheating
temperature, effects on the BBN becomes more significant.

\begin{figure}
    \centerline{\epsfxsize=0.7\textwidth\epsfbox{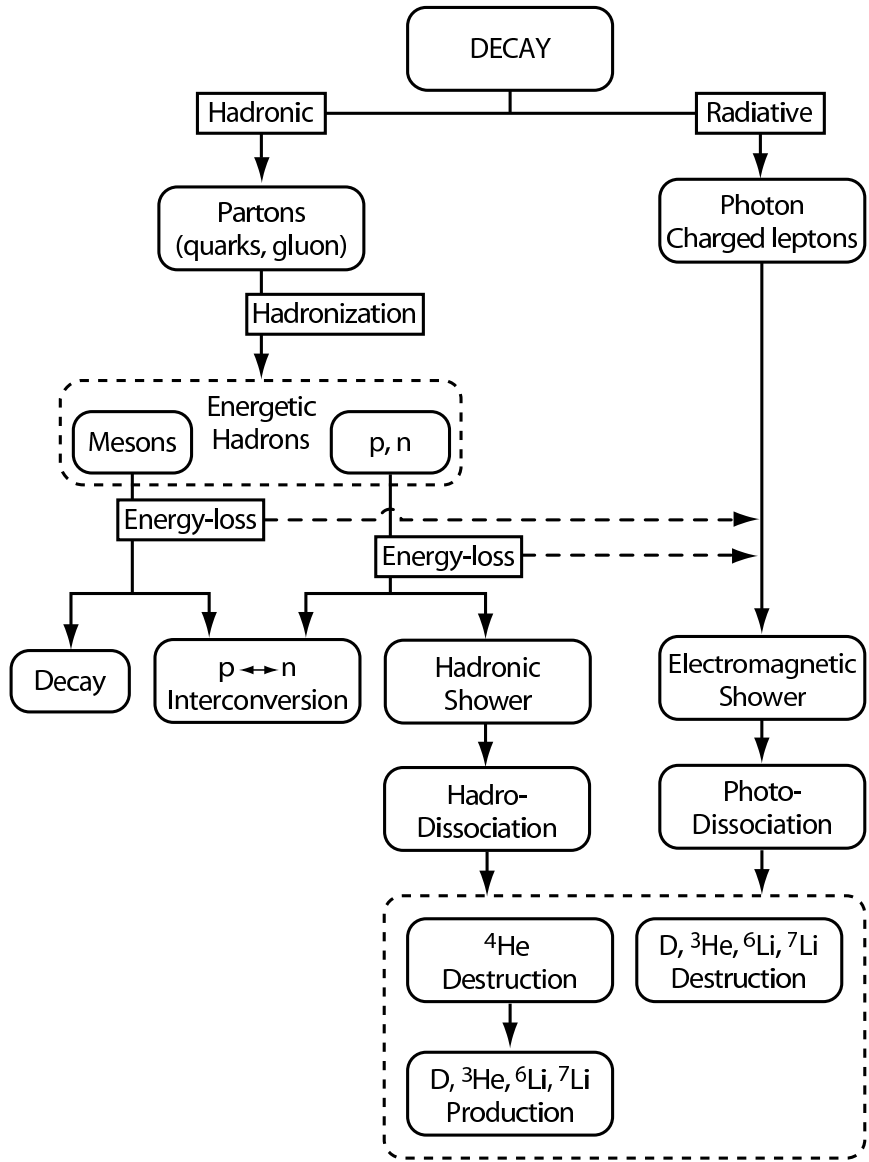}}
    \caption{Outline of our analysis.}
    \label{fig:hadronic}
\end{figure}

Once the primordial abundance of the gravitino is given, we consider
the effects of the decay of the gravitino on the BBN.  Importantly,
the interaction of the gravitino is well constrained by the local
supersymmetry and its lifetime is calculable.  (For the detailed
values, see the next section.)  Here, we assume several reasonable
values of the hadronic branching ratio $B_{\rm h}$ of the gravitino,
and calculated the abundances of the light elements taking account of
the effects of hadronic and electro-magnetic showers induced by the
decay products of the unstable gravitino.  Outline of our treatments
of the hadronic and electro-magnetic processes is schematically shown
in Fig.\ \ref{fig:hadronic}, and the details of our study are
discussed in the full papers\cite{Kawasaki:2004qu,KawMor(Koh)}.
(Notice that there are other recent studies on the BBN scenario with
late-decaying exotic particles\cite{RecentBBNhad}.)

We compare the theoretically predicted values of the light-element
abundances with the observations.  As observational constraints on the
light element abundances, we adopt the following values, D/H = $(2.8
\pm 0.4) \times 10^{-5}$\cite{Kirkman:2003uv}, $^4{\rm He}$ mass
fraction $Y_{\rm p} = 0.238 \pm 0.002 \pm 0.005 $ by Fields and Olive
(FO)\cite{Fields:1998} and $Y_{\rm p}= 0.242 \pm 0.002 (\pm
0.005)_{\rm syst}$ by Izotov and Thuan (IT)\cite{Izotov:2003xn},
$\log_{10}({\rm ^7Li}/{\rm H}) = -9.66 \pm 0.056 (\pm 0.3)_{\rm
syst}$\cite{Bonifacio:2002}, ${\rm ^6Li}/{\rm ^7Li}< 0.07
(2\sigma)$\cite{li6_obs}, and ${\rm ^3Hr}/{\rm D} < 1.13
(2\sigma)$\cite{Geiss93}.  The above errors are at 1$\sigma$ level
unless otherwise stated.  Then, we derive upper bound on the reheating
temperature requiring that the theoretical predictions be consistent
with the observations.

\section{Results}

Now, we show the results of our analysis.  Here, we consider two
typical case.  The first case is that the gravitino dominantly decays
into the gluino pair;  in this case, the hadronic branching ratio is
expected to be $1$ and, in addition, the lifetime of the gravitino is
estimated to be 
\begin{eqnarray}
    \tau_{3/2} (\psi_\mu\rightarrow g+\tilde{g}) \simeq 
    6 \times 10^7\ {\rm sec} \times
    \left( \frac{m_{3/2}}{100\ {\rm GeV}} \right)^{-3}.
\end{eqnarray}
The second case is that the gravitino decays into the photon and the
lightest neutralino (which we assume to be the photino); then, the
hadronic branching ratio is expected to be $\sim O(10^{-2}-10^{-3})$
and the lifetime is given by
\begin{eqnarray}
    \tau_{3/2}(\psi_\mu \rightarrow \gamma + \tilde{\gamma})
    \simeq 
    4\times 10^8{\rm ~sec}
    \left(\frac{m_{3/2}}{100 {\rm ~GeV}} \right)^{-3}.
\end{eqnarray}
For these two cases, we have derived the upper bound on the reheating
temperature.

\begin{figure}[t]
    \centerline{\epsfxsize=\textwidth\epsfbox{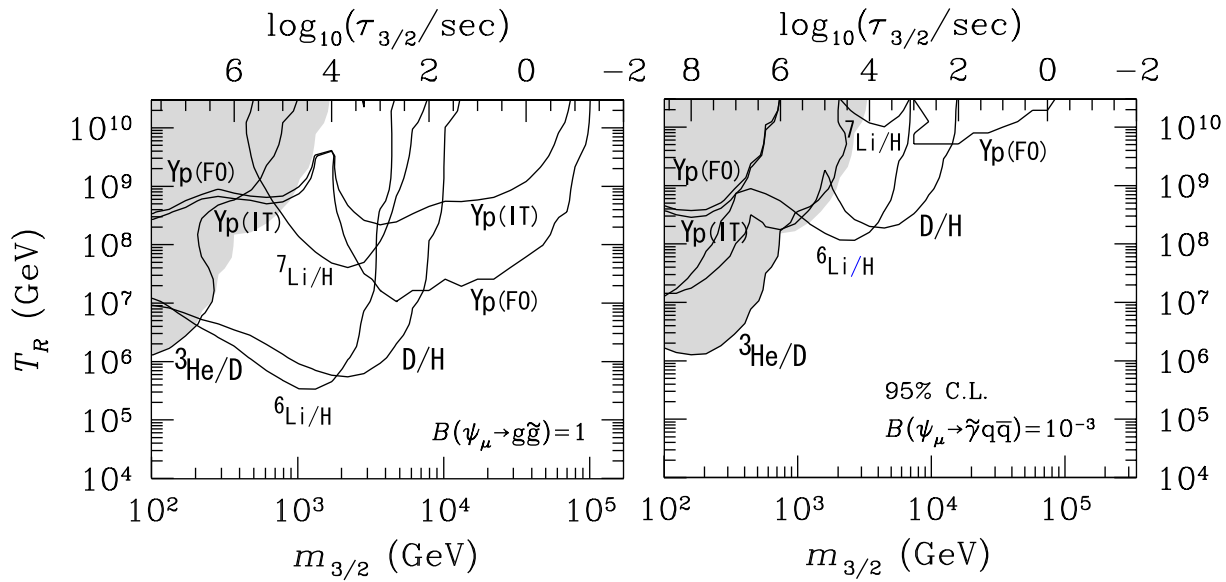}}
    \caption{Upper bound on the reheating temperature after the
    inflation.  The hadronic branching ratio of the gravitino is taken
    to be $1$ (left) and $10^{-3}$ (right).  The shaded regions are
    the excluded region for the case with $B_{\rm h}=0$.}
    \label{fig:TR}
\end{figure}

In Fig.\ \ref{fig:TR}, upper bound on the reheating temperature after
the inflation is plotted as a function of the gravitino mass.  The
results indicates that, as the hadronic branching ratio becomes
larger, constraint on the reheating temperature becomes severer.  For
the case with $B_{\rm h}=1$, for example, behavior of the upper bound
can be understood in the following way.  For $m_{3/2}\lesssim 200\
{\rm GeV}$, energetic neutron is likely to decay before scattering off
the background nuclei.  In this case, effects of the hadronic decay is
not so significant while, in this case, effects of the
photodissociation becomes comparable to or more significant than the
hadrodissociation.  Then, the strongest constraint is from the
overproduction of ${\rm ^3He}$; For $200\ {\rm GeV}\lesssim
m_{3/2}\lesssim 7\ {\rm TeV}$, energetic hadrons (in particular,
neutron) is hardly stopped by the electro-magnetic processes and hence
the hadrodissociation processes become the most efficient.  In
particular, in this case, non-thermal productions of ${\rm D}$ and
${\rm ^6Li}$ provide the most stringent constraint; For $7\ {\rm
TeV}\lesssim m_{3/2}\lesssim 100\ {\rm TeV}$, the $p\leftrightarrow n$
conversion processes are efficient and significant amount of $p$ may
be converted to $n$ resulting in the enhancement of $^4{\rm He}$.  In
this case, the constraint from the overproduction of ${\rm ^4He}$ is
the most significant; For $m_{3/2}\gtrsim 100\ {\rm TeV}$, gravitino
decays before the BBN starts.  In this case, no upper bound is
obtained on the reheating temperature.

\section{Summary}

In our study, we have studied the effects of the unstable gravitino on
the BBN, paying particular attention to the hadronic decay modes.  As
we have emphasized, as the hadronic branching ratio becomes larger,
constraints become more stringent.

Our results have significant implications.  In particular, for the
gravity-mediated supersymmetry breaking, gravitino mass is expected to
be $\sim O(100)\ {\rm GeV}$.  In this case, even if the hadronic
branching ratio is $\sim O(10^{-3})$, the reheating temperature is
constrained to be smaller than $10^6-10^8\ {\rm GeV}$.  If the
gravitino mass is much larger than $\sim O(100)\ {\rm GeV}$, the
constraint on $T_{\rm R}$ may be relaxed.  With such gravitino,
however, the hadronic branching ratio would be close to $1$ since, in
such a case, all the superpartners of the standard-model particles are
expected to be lighter than the gravitino from the naturalness point
of view.  (Such a mass spectrum may be realized in the
anomaly-mediated supersymmetry breaking scenario\cite{amsb}.)  For
$m_{3/2}\sim O(10-100)\ {\rm TeV}$ with $B_{\rm h}\sim 1$, the
upper bound is given by $T_{\rm R}\lesssim 10^7-10^{10}~{\rm GeV}$.
For the cosmology, since the reheating temperature is required to be
very low when the gravitino mass is $O(100\ {\rm GeV} - 1\ {\rm
TeV})$, baryogenesis should occur with very low reheating temperature.
This fact imposes significant constraints on some of the scenarios of
the baryogenesis, in particular for the leptogenesis scenario with
right-handed neutrinos\cite{Fukugita:1986hr}.

\section*{Acknowledgements}
This work is supported by the Grants-in Aid of the Ministry of
Education, Science, Sports, and Culture of Japan No.~14540245 ({MK}),
15-03605 ({KK}) and 15540247 ({TM}).

\end{document}